\documentclass[structabstract]{aa}  
\usepackage{graphicx}
\usepackage{savesym}
\usepackage{amsmath}
\usepackage{amssymb}
\usepackage{txfonts}
\usepackage{time}
\usepackage{natbib}
\bibpunct{(}{)}{;}{a}{}{,}
\usepackage[usenames]{color}

\begin{document}
\frenchspacing

\title{Stokes imaging polarimetry using image restoration:\\A calibration strategy for Fabry-P\'{e}rot based instruments}
\titlerunning{Stokes imaging polarimetry using image restoration: A calibration strategy for Fabry-P\'{e}rot based instruments}

\author{R.\ S.\ Schnerr\inst{1,2}
  \and J.\ de la Cruz Rodr\'iguez\inst{1,2}
  \and M.\ van Noort\inst{1,2,3}
}
\authorrunning{Schnerr et al.}

\offprints{R.S.S.: \email{roald@schnerr.nl}}

\institute{Institute for Solar Physics of the Royal Swedish
  Academy of Sciences, AlbaNova University Center,
  SE-106\,91 Stockholm, Sweden
  \and
  Department of Astronomy,
  Stockholm University,
  AlbaNova University Center, SE-106\,91 Stockholm, Sweden
  \and
  Max-Planck-Institut f\"{u}r Sonnensystemforschung, Max-Planck-Str. 2, 37191 Katlenburg-Lindau, Germany
}

\abstract
{The combination of image restoration and a Fabry-P\'{e}rot interferometer (FPI) based instrument in solar observations results in specific calibration issues. FPIs generally show variations over the field-of-view, while in the image restoration process, the 1-to-1 relation between pixel space and image space is lost, thus complicating any correcting for such variations.}
{We develop a data reduction method that takes these issues into account and minimizes the resulting errors.}
{By accounting for the time variations in the telescope's Mueller matrix and using separate calibration data optimized for the wavefront sensing in the MOMFBD image restoration process and for the final deconvolution of the data, we have removed most of the calibration artifacts from the resulting data.}
{Using this method to reduce full Stokes data from CRISP at the SST, we find that it drastically reduces the instrumental and image restoration artifacts resulting from cavity errors, reflectivity variations, and the polarization dependence of flatfields. The results allow for useful scientific interpretation. Inversions of restored data from the $\delta$ sunspot AR11029 using the Nicole inversion code, reveal strong ($\sim$10 km\,s$^{-1}$) downflows near the disk center side of the umbra.}
{The use of image restoration in combination with an FPI-based instrument leads to complications in the calibrations and intrinsic limitations to the accuracy that can be achieved. We find that for CRISP, the resulting errors can be kept mostly below the polarimetric accuracy of $\sim$10$^{-3}$.
Similar instruments aiming for higher polarimetric and high spectroscopic accuracy, will, however, need to take these problems into account.}
\maketitle

\section{Introduction}
As technical possibilities increase, more advanced and complicated instruments are being designed for observational studies of the Sun.
New instrumentation is generally designed to offer better spatial resolution, better spectral coverage, and full polarization information. Higher spatial resolution will help resolve, e.g., small-scale magnetic fields in the quiet Sun. In sunspots, the magnetic fields and flows in penumbra, lightbridges, and umbral dots make interesting targets. Better spectral coverage and polarization information are needed to enable meaningful interpretation of the observed features. Adaptive optics (AO) are currently available at the major ground-based solar observatories and have greatly improved the performance of telescopes.

Additionally, postprocessing of high-cadence imaging data to correct the blurring effect of seeing has made a huge impact on image quality. Image restoration methods such as MOMFBD \citep[Multi-Object Multi-Frame Blind Deconvolution,][]{vannoort:2005,lofdahl:2002} and Speckle interferometry \citep[see][and references therein]{vonderluhe:1993} have become essential for ground-based, high spatial-resolution solar imaging. Because these methods allow combining all the images obtained within a dataset, while retaining high image quality, it removes the need for frame selection and results in more reliable and higher signal-to-noise data.

In this context, Fabry-P\'{e}rot interferometer (FPI) based instruments have become quite popular and are in operation at many solar observatories,  e.g.\ CRISP \citep{scharmer:2008} at the Swedish 1-m Solar Telescope (SST), the Telecentric etalon solar spectrometer \citep[TESOS,][]{kantischer:1998}, and 
G\"ottingen Fabry-P\'erot spectropolarimeter \citep{Puschmann:2006} at the Vacuum Tower Telescope and the Interferometric bidirectional spectrometer \citep[IBIS,][]{cavallini:2006} at the Dunn Solar Telescope. Similar instruments are likely to be developed for the future 4-m class solar telescopes ATST and EST. The main advantages of FPIs over Lyot filters are that they have a high transmission, allow for rapid wavelength tuning, and allow for a dual-beam polarimetric setup with a polarizing beam splitter close to the final focal plane. A disadvantage is that even very well-made FPIs will show variations in the width and central wavelength of the transmission peak over the field of view \citep[FOV, e.g.][]{ibis:2008}. When the FPIs are set up in a telecentric mount, as with CRISP, these variations cause intensity variations on relatively small spatial scales.
In collimated mount, such small-scale features disappear, but the central wavelength shifts systematically away from the center of the image. 
For further comments on FPIs in telecentric vs.\ collimated mount see \citet{scharmer_crisp:2006}.

We plan to discuss a theoretical approach to the processing of full-Stokes line scans from FPI-based instruments in a future paper, but here we present the practical application of that approach to real CRISP data of a sunspot observed in the 6302.5 \AA\ \ion{Fe}{i} line, including all the necessary data reduction steps, such as correction for telescope polarization, flatfielding, image restoration, and demodulation.
In Sect.\ \ref{sect:pol_acq} we discuss the general setup of the data acquisition procedures, and in Sect.~\ref{sect:instr_calib} the challenges and possibilities for calibration procedures.
The proposed calibration procedure is presented in Sect.~\ref{sect:cal_scheme}, results of the procedure applied to CRISP data are presented in Sect.~\ref{sect:results}, and the conclusions in Sect.~\ref{sect:conclusions}.

\section{Polarimetric data acquisition}
\label{sect:pol_acq}

\subsection{The optical setup}
\begin{figure}[tbhp]
\centering
\includegraphics[width=0.8\columnwidth, trim=20 0 20 0]{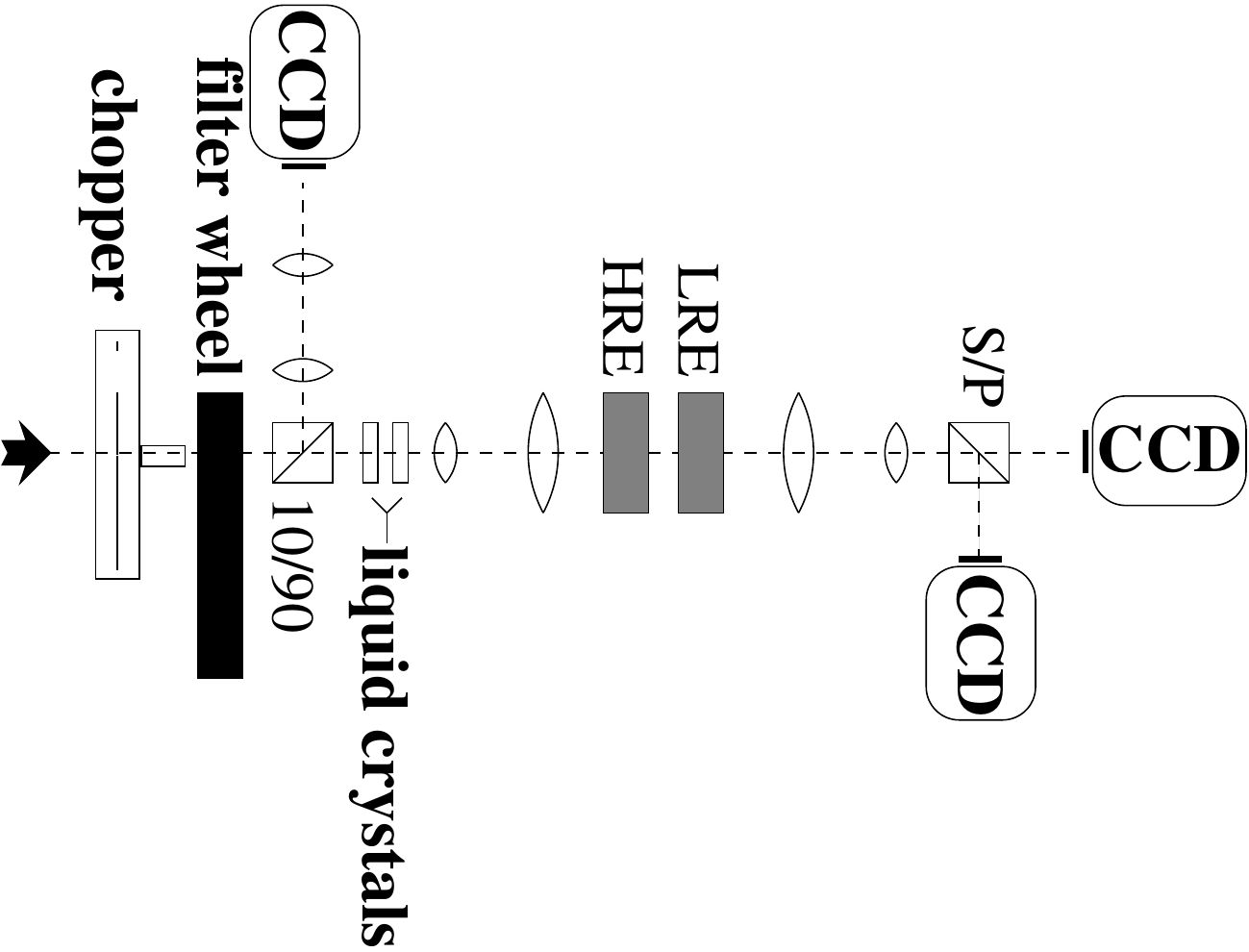}
\caption{Schematic drawing of the optical set-up at the SST. After reflecting off the dichroic mirror separating the red and blue beams (not shown), the red beam passes the chopper and prefilter. Then a beam splitter sends 10\% of the light to the WB camera. The other beam passes the LCs, the high- and low-resolution etalons of the CRISP filter (HRE and LRE resp.), and finally a polarizing beam splitter separating the p- and s-polarized light to the transmitted and reflected beams. All three cameras are 1024$\times$1024 pixel CCD (Sarnoff) cameras running at $\sim$36 Hz.}
\label{fig:optsetup}
\end{figure}

The data acquisition setup used with CRISP at the SST (see Fig.~\ref{fig:optsetup})
is similar to the one used by \citet[][]{2008A&A...489..429V}, with the notable difference that 
the SOUP Lyot tunable filter has been replaced with the CRISP Fabry-P\'{e}rot filter 
\citep{scharmer:2008}.
The significantly higher ($\sim8\times$) peak transmission of the Fabry-P\'erot filter results in a higher S/N for similar exposure times than with the Lyot filter, and the use of a dual beam polarimetric setup results in a significant reduction of the seeing-induced crosstalk.
In addition, the higher tuning speed of CRISP allows for observing programs scanning over $\sim$10 spectral points instead of the two to four points one would typically get with the Lyot filter.
A downside, however, is that the placement of the filter cavities in a telecentric mount 
configuration leads to small-scale pass-band shifts and transmission profile width variations across the FOV, due to
cavity and reflectivity errors. In the case of CRISP, the high- and low-resolution FPIs have been placed $\sim$10 cm away from the focal plane in order to smooth out such variations.

CRISP data are obtained while continuously cycling through a 4-state liquid-crystal (LC) scheme, which converts combinations of the incoming polarization states Q, U, and V to linear polarization that can be analyzed with the polarizing beam splitter close to the final focal plane. Data are simultaneously recorded using two narrow-band cameras (called transmitted and reflected camera) in a dual-beam setup and a wide-band camera (receiving 10\% of the light passing the prefilter via a beam-splitter) placed before the liquid-crystal variable retarders. The frame rate of $\sim$36 Hz is set by the rotating chopper, which has a duty cycle setting the exposure time at 16 ms. CCD readout is performed during the dark part of the cycle. For the data shown in this paper, 4 full LC-cycles were completed before tuning to the next of a total of 12 wavelengths. The complete linescan of 4 LC-states $\times$ 4 repeats $\times$ 12 wavelengths = 202 frames covers $\sim 6$ seconds.

\subsection{Calibration data}
The data required for calibrating the science data from CRISP are obtained in the following way.
The dark images are recorded in a similar way as the science data, but with the light path blocked after the exit window of the vacuum tube.
The final dark is the average of a large number of darks (typically 100--1000). Flatfields are obtained by summing many ($\sim$1000) images of the Sun taken with the telescope pointing moving across the solar disk. As such, the flatfields include both transmission variations over the FOV as actual CCD gain variations, as do the gains that are derived from these flatfields. Therefore, in this paper, ``flatfielding`` and ``gain correction`` imply the same correction.
The polarization calibration data is obtained by producing light with known polarization states using a linear polarizer and a quarter-wave plate placed below the exit window of the vacuum tube in the tower. The modulation scheme is then calibrated using the method described in Appendix~\ref{appa}. The (time-dependent) telescope Mueller matrix has been measured by \citet{jakob:2006} and is assumed to be known here.

\subsection{Image restoration}
For ground-based, high-resolution imaging, image restoration is a vital tool. However, it is especially important for multi-wavelength and polarimetric observations, because it will reduce differential image shifts and blurring between images recorded at different wavelengths and polarization states. For polarimetric data, this implies that the image restoration has to be done {\it before} demodulating the data (to convert the LC-states to Stokes parameters I,Q,U,V), because the images of the different LC-states are each associated with a particular seeing disturbance that must be compensated for individually, to reduce artificial, seeing-induced signals.
In our analysis we focused on image restoration using MOMFBD \citep{vannoort:2005}, but some of the complications discussed below are equally relevant for Speckle restorations.

MOMFBD has several properties that can complicate calibrations.
\begin{enumerate}
 \item The direct relation between an image pixel in the original data and the restored image is lost.
 \item Artificial features in images can influence the wavefront sensing and can be amplified by the deconvolution.
 \item Input images are assumed to be monochromatic, although MOMFBD can handle different wavelength channels within one restoration.
 \item Noise is assumed to follow a Gaussian distribution with a constant width over each isoplanatic patch, which can be a problem when high-contrast targets are observed.
\end{enumerate}

The loss of the direct relation between input data pixels and restored data pixels means that all calibrations that cannot be done {\it before} image restoration, and hence before demodulating, cannot be done on a pixel-by-pixel basis, but involve the point-spread-functions (PSFs, which result from seeing and the finite pupil size) of the input images. Similarly, artificial features in the input images should be removed as much as possible before image restoration, because they will interfere with the wavefront sensing and can be strongly amplified by the deconvolution. Input images that are not monochromatic over the FOV are rather hard to deal with, because this cannot be properly corrected without knowing the entire line profile. 

Both the wavelength changes over the FOV and the assumption of noise with a Gaussian distribution of a constant width are limiting factors for the maximum (polarimetric) accuracy that can be obtained when using image restoration.
To avoid interference from artifacts in the narrow band cameras on the wavefront sensing, one can only use images from the wide band camera for the wavefront sensing and apply the corrections to the data of all cameras. 
However, the additional wavefront information contained in the narrow band data is then not used for the image restoration.

The changing beam configuration due to seeing corrections by the AO (and tip-tilt mirror) in principle also affects data calibration, because it changes the way the optics in front of the AO are sampled. However, it does not prevent pixel-based calibrations, and since we assume the telescope matrix is constant over the FOV, the effects of the changing beam configuration are not considered here.
Image restoration of the data used in this paper was performed with MOMFBD assuming isoplanatic patches of 96$\times$96 pixels ($\sim$5.7$\times$5.7\arcsec) and using 37 Karhunen-Lo\`eve modes for modeling the wavefront phases.

\section{Instrument calibration}
\label{sect:instr_calib}

A straightforward calibration scheme typically involves the following steps.
\begin{enumerate}
 \item Dark-correct data from the different cameras (transmitted, reflected, and wide band).
 \item Gain-correct the different LC-state and wavelength images of all cameras.
 \item Use image restoration to combine the data for the different cameras and states, and correct for the seeing.
 \item Demodulate transmitted and reflected beams using the polarization calibration matrix (see Appendix~\ref{appa}).
 \item Combine Stokes images from the transmitted and reflected beams to reduce seeing crosstalk.
 \item Compensate for telescope polarization using the inverse of the telescope Mueller matrix.
\end{enumerate}

Due to specific instrumental properties, such as polarization-dependent flatfields and FPI inhomogeneities, polarimetric and spectral calibration problems arise in steps 2 to 4.
These are discussed in more detail below.

\subsection{Polarimetric issues}
For polarimetric calibration of the data it is convenient to split the calibration in two parts.
\begin{enumerate}
 \item {\bf Telescope}, from the 1-m primary lens at the entrance of the vacuum tube down to the exit window at the bottom of the vacuum tube. Due to the 45$^\circ$ angle of incidence on the rotating azimuth- and elevation-mirrors, this part is intrinsically time variable, but is assumed to be independent of the line-of-sight (constant over the FOV). A model describing the Mueller matrix of the telescope as a function of the azimuth and elevation angles at 6302 \AA\ was determined by \citet{jakob:2006}.
 \item {\bf Instrument}, from the vacuum exit window to the CCDs. This part is FOV-dependent (e.g.\ dust close to the focal plane), but can be calibrated daily using calibration optics and is assumed not to vary in time. It is described by a modulation matrix, which converts the input Stokes vector into intensities measured with the LC-states. How this matrix can be determined is described in Appendix~\ref{appa}.
\end{enumerate}

\noindent We can describe the total conversion matrix of the system (${\cal A}$) as a matrix product of the Mueller matrix of the telescope ${\cal T}$ and the modulation matrix of the instrument ${\cal M}$:
\begin{equation}
{\cal A}(x,y,t) = {\cal M} (x,y) \cdot {\cal T} (t),
\end{equation}
where the transmitted and the reflected beams have their own (although related) modulation matrices. Because the intensity of the light during the determination of the telescope matrix and the modulation matrix is not known, an additional gain calibration is still required.
The telescope matrix ${\cal{T}}(t)$ varies with time, so the total system response matrix ${\cal A}$ is time dependent.
This means that unless the flatfield images and the data are recorded at the same 
time, which is not practical, an appropriate gain correction (for the modulated data) must be computed. To this end, the flatfield must first be
transformed to a time-independent location. The obvious (and only) place available for this is the plane of the sky, i.e.\ on the Sun, since we assume that all time variability is contained in the telescope matrix. The details of this procedure are discussed in Sect.~\ref{sect:cal_scheme}.

\subsection{Spectral issues}
An important calibration problem is that part of the intensity changes over the field of view are not the result of real CCD gain
variations, but of variations in FPI reflectivity and cavity size \citep[see, e.g.,][]{ibis:2008,cauzzi:2009}. When observing at a wavelength where $\delta I/\delta \lambda \neq 0$, e.g.\ in the wing of a 
spectral line, a wavelength shift due to a cavity error results in a change in the observed intensity that is completely unrelated to the 
gain of that particular pixel. Similarly, variations in the reflectivity cause changes in the integrated transmission profile.

The question, however, is which flatfield correction should be applied to the data. Although the intensity variations due to cavity errors are not real gain changes, the same cavity errors also affect the science data. The line profiles in the science data can be very different, though, so the intensity changes will in general not be the same as in the flatfield data. A change in wavelength scale due to a cavity error just {\it cannot} be corrected in a flatfielding procedure, which only changes the intensity, unless one knows the exact line profile. For a data calibration scheme without image restoration, one could attempt to somehow determine the real pixel gains, apply those, and account for the wavelength shifts later on in the analysis. For image restoration, however, the intensity variations that then remain could pose a problem for the wavefront sensing and/or the deconvolution.

The strategy that we propose is discussed in the next section.

\section{Calibration scheme}
\label{sect:cal_scheme}
For the wavefront sensing part of the image restoration, it is important to use images that are as
monochromatic as possible and that have no sharp artifacts. 
Sharp features can result from CCD pixel rows that have different responses to the polarization of the incoming Stokes vector, when the S/P beamsplitter (see Fig.~\ref{fig:optsetup}) is not 100\% effective.
These pixels then stand out in some of the matrix elements of ${\cal M}$.
Such features should be reproduced in flatfields that are properly converted to the time of the
observations, because this information is contained in the modulation matrix, as long as the polarization of the
incoming light is low, the telescope polarization is known, and the instrument is stable. Therefore these features should be removed from the data by flatfielding.

Dealing with the non monochromatic nature of the data, resulting from FPI transmission variations over the FOV, is more complicated. In the final restored data, one can attempt to take the wavelength shifts caused by cavity errors into account, although as a result of the image restoration process the effective shift in a given pixel in the restored image is always some mixture of the shifts of all CCD pixels within the PSFs of the dataset. For the image restoration, the intensity changes are more problematic, especially if they contain high spatial frequencies, because these distort the wavefront sensing. Intensity changes due to reflectivity and cavity errors of the FPI cannot be interpreted as simply brighter or darker regions in the observed object by the wavefront sensing, because they do not move with the object while the seeing varies throughout a dataset.

Our approach is to use flatfields that make the data as smooth as possible, before using the data for the wavefront sensing. This is done using the recipe described in Sect.~\ref{sect:momfb_flats}. These flatfields {\it include} intensity changes due to varying cavity properties over the FOV. This is a straightforward choice, but a relatively arbitrary one, because such a flatfield will only properly correct these intensity changes if the observed line profile is the same as the one of the flatfield data, which is in general not the case.
For the final restored data however, we do not want to use these flatfields, because they modify the observed line profiles, but rather those flatfields that are properly corrected for the varying cavity properties.

A relatively simple way to obtain flats that are not affected by cavity errors in the FPI would be to use a wavelength close to the observed line where $\delta I/\delta \lambda =0$.
However, such a method does not account for the wavelength dependence of the flatfields from variations in the prefilter transmission profile, LRE/HRE co-tuning errors and the wavelength dependent fringes that are observed in the Sarnoff cameras. Additionally, the wavelength where such flats have to be taken might not correspond to the preferred continuum point, because the instrumental response has to be taken into account, increasing the time required for calibrations.
Therefore we developed a more accurate routine to derive proper flatfields, which simultaneously solves for the FPI cavity and reflectivity errors, and prefilter changes over the FOV. This routine is described in Sect.~\ref{sect:real_gains}.

\subsection{Time-independent and image-restoration flatfields}
\label{sect:momfb_flats}
The proposed scheme thus consists of creating flats especially designed to minimize image restoration artifacts and time-independent flatfields that should properly flatfield the data without modifying the line profile. The image restoration flatfields are applied before the image restoration process, but removed again afterwards when the time-independent flatfields are applied.

\begin{figure*}[tbhp]
\centering
\includegraphics[width=\linewidth, trim=0 0 0 0]{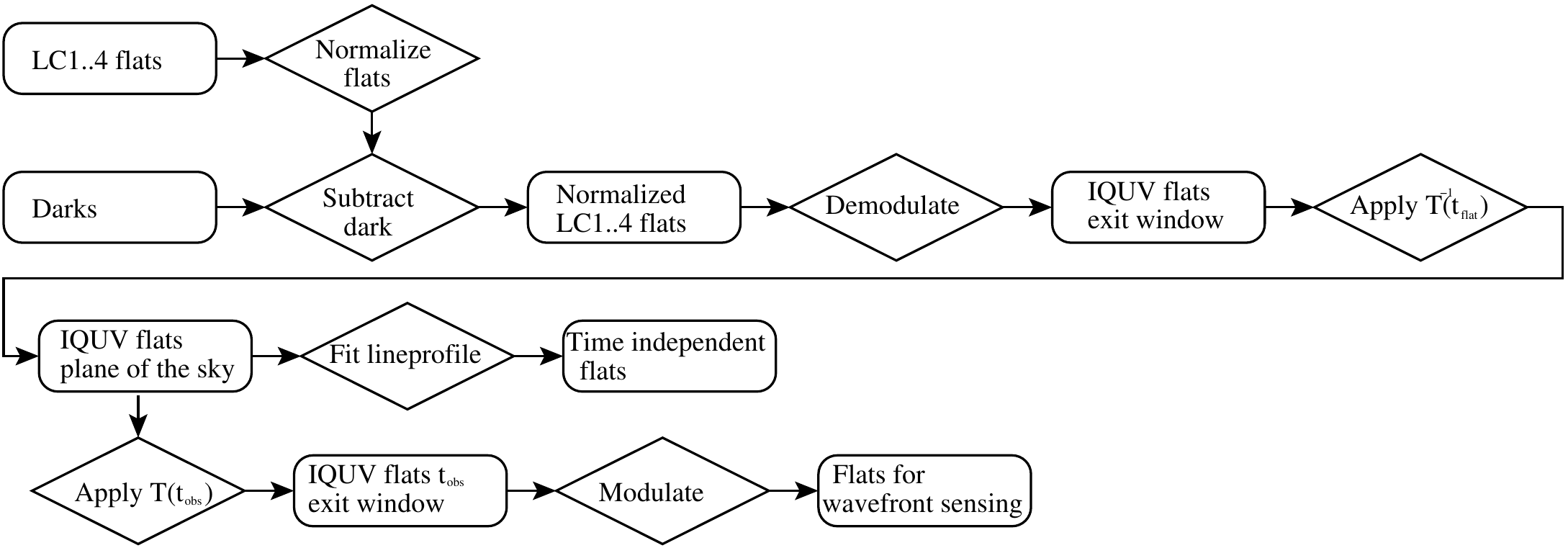}
\caption{Flowchart illustrating the procedure used to calculate the flatfields used for wavefront sensing and the time-independent flatfields corrected for wavelength shifts.}
\label{fig:flowchart1}
\end{figure*}

To obtain time-independent flatfields, we have to invert the modulation matrix of each pixel to obtain the demodulation matrix (${\cal M}^{-1}$, the inverse of the modulation matrix ${\cal M}$).
By applying the demodulation matrix to the flatfield data of the 4 LC-states, we obtain flatfield images for all Stokes parameters under the vacuum-tube's exit window.
Using the inverse Mueller matrix of the telescope, valid for the time at which the flatfields were taken, we can calculate the Stokes 
flats on the Sun (on the plane of the sky). The flatfields represent the effective flatfields for Stokes I, Q, U, and V. An overview of how both the time-independent and image-restoration flatfields are determined is shown in Fig.~\ref{fig:flowchart1}.

To calculate the flatfields relevant for the time the observations were obtained, one has to apply
the telescope matrix applicable for that time to the flatfields on the Sun and then apply the modulation matrix. For this we only use the Stokes I flatfield (and set Stokes Q, U, and V to 0) to avoid imprinting a polarization pattern on our LC1..4 flats (which results from the FOV-variations in the Stokes QUV-to-LC1..4 components of ${\cal M}$). These flats will include intensity changes due to cavity and reflectivity errors, which should smooth these variations in the data (when wavelength shifts due to line-of-sight velocity variations are small compared to the cavity errors). Also the polarization levels should be properly reproduced to the accuracy that the telescope Mueller matrix and modulation matrix are known, if the intrinsic polarization of the observed target is low.

\subsection{Determining the proper flatfields}
\label{sect:real_gains}
The FPI has small variations in the cavity size and reflectivity, which results in changes in the central wavelength and width of the transmission profile, 
respectively. In addition, the central wavelength of the prefilter shows a slight variation over the FOV as the angle of incidence on the prefilter changes (due to the physical dimensions of the CCD).
To determine the real transmission over the FOV, one has to account for these effects. At the SST, flatfields are determined by summing many ($\sim$1000)
exposures per state and wavelength obtained while the telescope pointing is moving across the solar disk (to avoid intensity changes over the FOV due to limb darkening). If one assumes that
the observed line profile of the averaged flatfield is constant over the FOV, then the different factors can be determined by using an iterative fitting scheme.

The transmission of a single FPI is given by the Airy function \citep[e.g.][]{kantischer:1998}:
\begin{equation}
T_\mathrm{FPI} = \frac{1}{1+F \sin^2 (2\pi d n \cos \theta /\lambda)}
\end{equation}
where the finesse $F$ is given by
\begin{equation}
F = \frac{4R}{(1-R)^2}
\end{equation}
with $R$ the reflectivity, $d$ the cavity size, $n$ the refractive index in the cavity, $\theta$ the angle of incidence, and $\lambda$ the wavelength. The total transmission $\Psi$ of the dual FPI setup of CRISP is calculated by multiplying the transmission of the high-resolution ($T_\mathrm{hr}$) and low-resolution FPI ($T_\mathrm{lr}$), assuming $\theta=0$ and assuming that cavity and reflectivity errors in the low-resolution FPI can be neglected because those in the high-resolution FPI will be dominant:
\begin{equation}
 \Psi(R,\Delta R) = T_\mathrm{lr}(R) \cdot T_\mathrm{hr}(R+\Delta R).
\end{equation}
Because the spectral range used is only a small fraction of the prefilter transmission profile width, we can limit the number of free parameters by not including the average prefilter shape in our fits, but instead only fit the deviations from the mean prefilter shape, assuming a linear dependence on wavelength.

Given the constant line profile $\phi(\lambda)$, average transmission profile $\Psi(R)$ of the combined FPIs, and (wavelength independent) gain $g$, the observed line profile in pixel $i$ will be given by
\begin{equation}
 \phi_{obs,i}(\lambda) = g_i \cdot [(1+D_{pf,i}) \cdot (\lambda-\lambda_0)] \cdot \Psi(R+\Delta R_i) * \phi(\lambda - \Delta \lambda_i)
\label{eq:lineprof_fit}
\end{equation}
where $g_i$ is the pixel gain, $D_{pf,i}$ the prefilter deviation factor, $\Delta R_i$ the reflectivity error, $\Delta \lambda_i$ the cavity error, and $\lambda_0$ the central wavelength of the line.

\begin{figure}[tbhp]
\includegraphics[trim=0 0.65cm 0 0,clip,width=0.99\columnwidth]{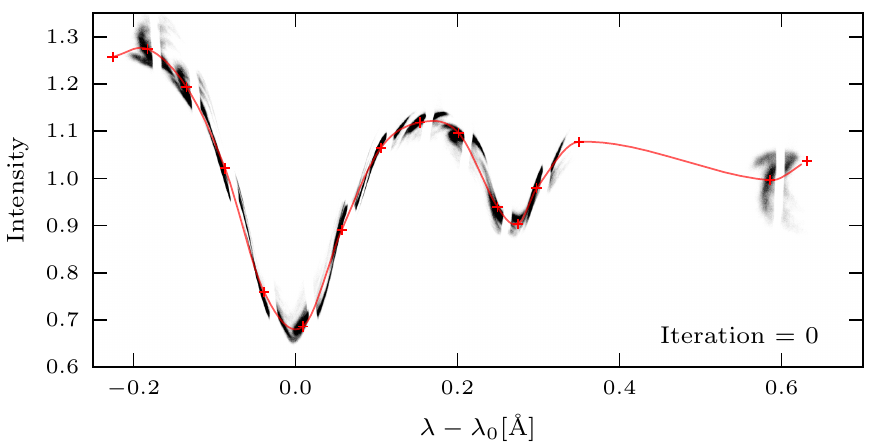}
\includegraphics[trim=0 0 0 0,clip,width=0.99\columnwidth]{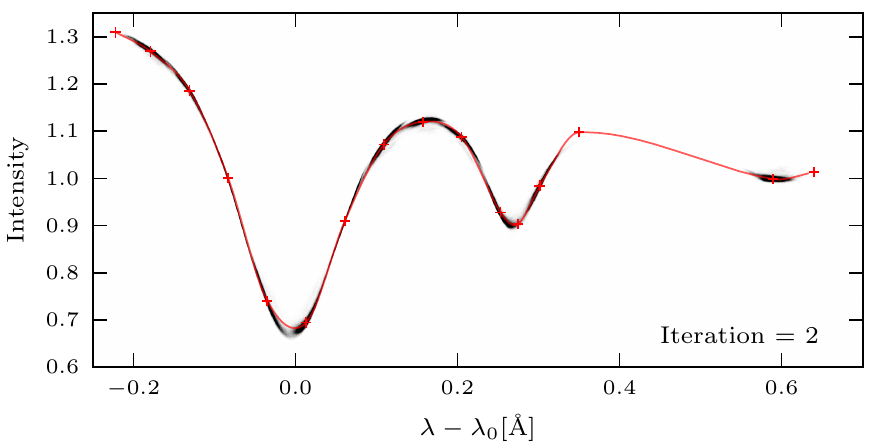}
\caption{Density plot of intensity (arbitrary units) vs.\ wavelength relative to line core for all points in the summed flatfield data before the first iteration (\textit{top}, where the line center has been determined using a parabolic fit to the line core) and the same plot after the second iteration (\textit{bottom}). The average line profile $\phi(\lambda)$, described by a spline curve through several nodes (red plusses), is overplotted in red.}
\label{fig:spline_interpolation}
\end{figure}

The iterative scheme is initiated by setting $g_i$ to the inverse of the mean intensity of the line profile, $\lambda_0 - \Delta \lambda_i$ to the minimum of a parabolic fit to the line minimum, and $D_{pf,i}$ and $\Delta R_i$ to zero. After applying these corrections, one can overplot all the observed line profiles and determine the average line profile by calculating a spline through a set of points at fixed wavelengths, for which the intensity is determined by a $\chi^2$ minimalization routine \citep[see Fig.~\ref{fig:spline_interpolation}; all fitting was done using the mpfit package, see][]{markwardt:2009,more:1978}. The ''real`` line profile is then calculated by deconvolving this fitted line profile with the average transmission profile.
As we now have an estimate for the average line profile and transmission profile, we can fit for the four unknown parameters in Eq.~\ref{eq:lineprof_fit} for each pixel. These parameters allow us to improve our estimate for the average line profile and transmission profile. These two steps can be repeated until no more improvement is made.

\begin{figure*}[tb!hp]
 \includegraphics[width=0.33\linewidth, trim=5 0 5 0, clip]{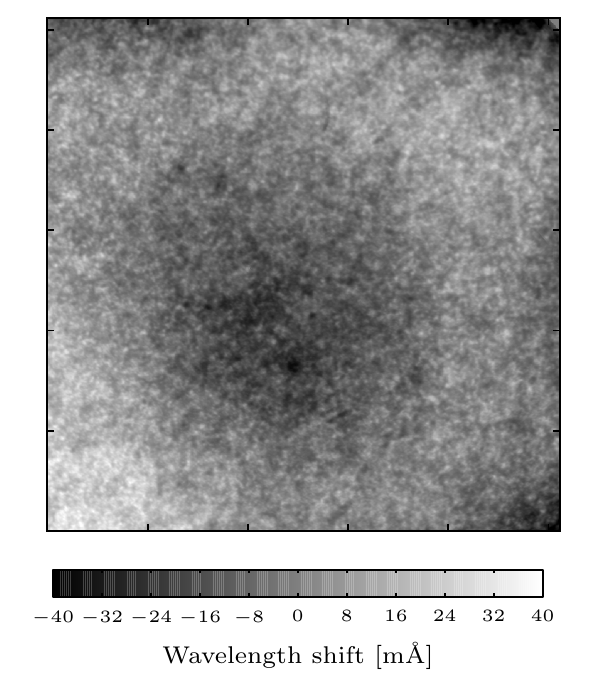}
 \includegraphics[width=0.33\linewidth, trim=5 0 5 0, clip]{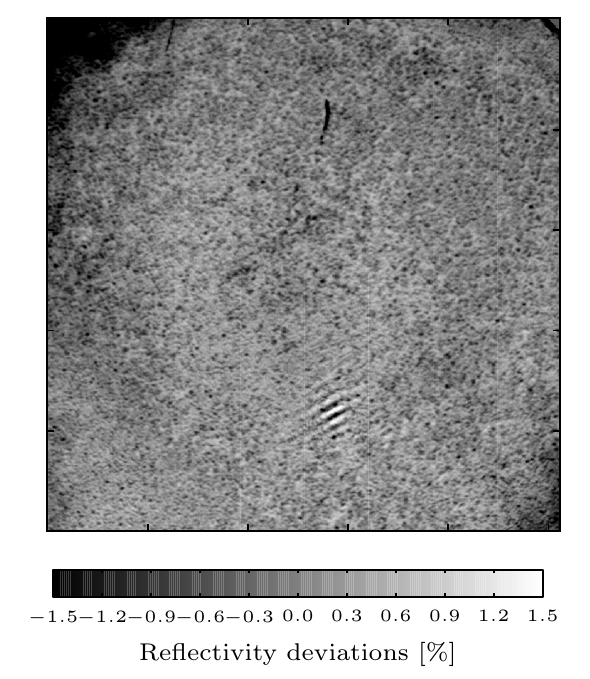}
 \includegraphics[width=0.33\linewidth, trim=5 0 5 0, clip]{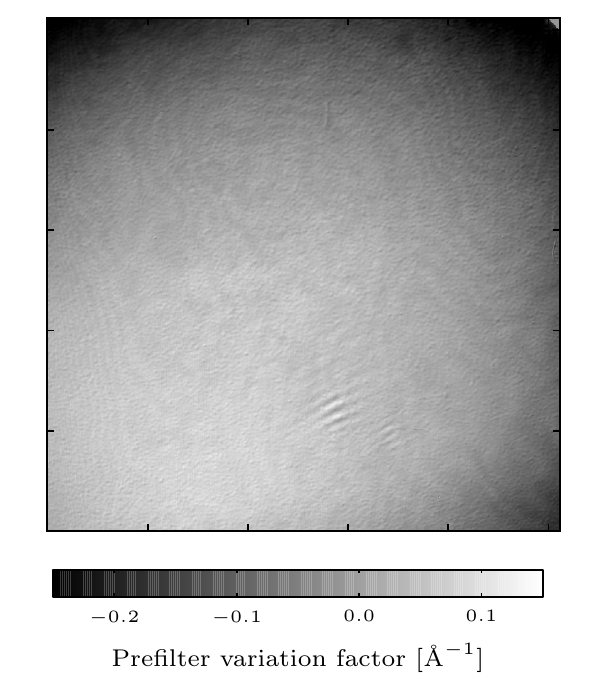}
 \caption{Wavelength shift resulting from cavity errors ({\it left}), variations in the FPI reflectivity ({\it middle}, nominal value 93.5\%), and the prefilter variation factor ({\it right}) as determined by the fitting routine described in Sect.~\ref{sect:real_gains} (see also Eq.~\ref{eq:lineprof_fit}) when applied to the 6302 \AA\ dataset presented in Sect.~\ref{sect:results}.}
 \label{fig:fitting_params}
\end{figure*}

Results of this fitting routine applied to a real dataset (see Sect.~\ref{sect:results}) are shown in Figs.~\ref{fig:spline_interpolation} and \ref{fig:fitting_params}. The total gain of a pixel is determined by the wavelength-independent part $g_i$ and a wavelength-dependent part given by the ratio of the observed intensity and the fitted line profile, as shown in Fig.~\ref{fig:spline_interpolation}.

\subsection{The prefilter transmission profile}
After the prefilter deviation factor and average line profile have been determined, we use the average line profile of all the pixels (corrected for cavity errors and gain) to determine the average prefilter shape $T_\mathrm{PF}$. The prefilter, which is mounted at a slight angle to tune the central wavelength, is assumed to have a Lorentzian profile:
\begin{equation}
 T_\mathrm{PF}(\lambda) = \frac{1}{1+(2\,(\lambda - \lambda_\mathrm{0,PF})/w)^{2N_\mathrm{cav}}}
\end{equation}
where $\lambda_\mathrm{0,PF}$ is the prefilter central wavelength, $w$ the full-width at half-maximum (FWHM), and $N_\mathrm{cav}$(=2) the number of cavities of the prefilter.

\begin{figure}[tbhp]
 \includegraphics[width=0.99\linewidth, trim=0 0 0 0, clip]{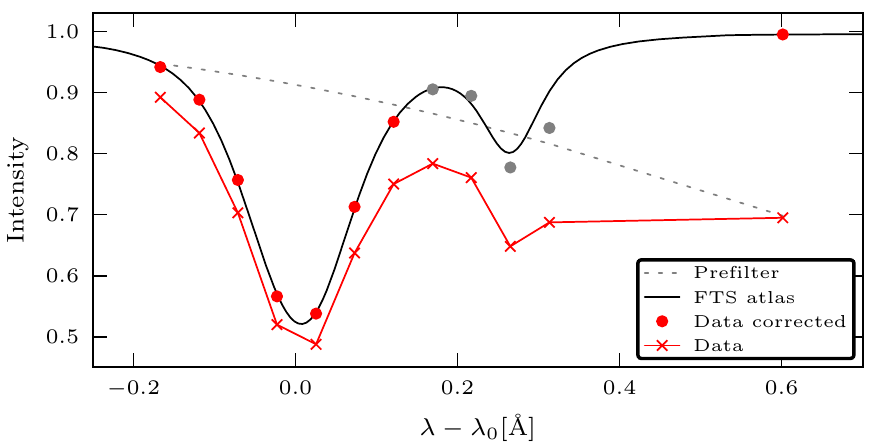}
 \caption{The prefilter shape (dashed line) is determined by matching the uncorrected average spectrum (crosses) to the FTS convolved with the CRISP transmission (full black line). After correcting for the prefilter, an excellent match is found (red circles). The points in the telluric line (gray circles) were not used in the procedure.}
 \label{fig:prefilter_fit}
\end{figure}

The prefilter parameters $\lambda_\mathrm{0,PF}$ and $w$ are determined by matching the average observed line profile to the line profile of the FTS atlas \citep[the atlas acquired with the Fourier transform spectrometer at the McMath-Pierce Telescope, available from][]{fts_atlas} convolved with the average transmission of CRISP (see Fig.~\ref{fig:prefilter_fit}), masking the points in the telluric line.
We find a prefilter width of 4.7$\pm$0.4~\AA, which agrees with the specification of 4.4-4.6~\AA, and a central wavelength of 6301.2$\pm$0.2~\AA.

\subsection{Iterative versus direct method}
After calculating both the actual CCD gains that we want to apply in the end and flatfields that should be used for the image restoration, we still have to account for the
impact of the image restoration process, as the result of which the 1-to-1 pixel relation is lost. Here we discuss two possible approaches to that problem:

\begin{itemize}
 \item {\bf Direct method}. Apply the flatfields prepared for the image restoration to the data and after the image restoration remove these flatfields and apply the time-independent flatfields, taking the average of the PSFs determined by the image restoration process into account.
 \item {\bf Iterative method}. Use the PSFs that have been determined by the image restoration to iteratively determine the Stokes images that best reproduce the observed data given the telescope Mueller matrix, modulation matrix, PSFs, and gains.
\end{itemize}

\begin{figure*}[tbhp]
\centering
\includegraphics[width=\linewidth, trim=0 0 0 0]{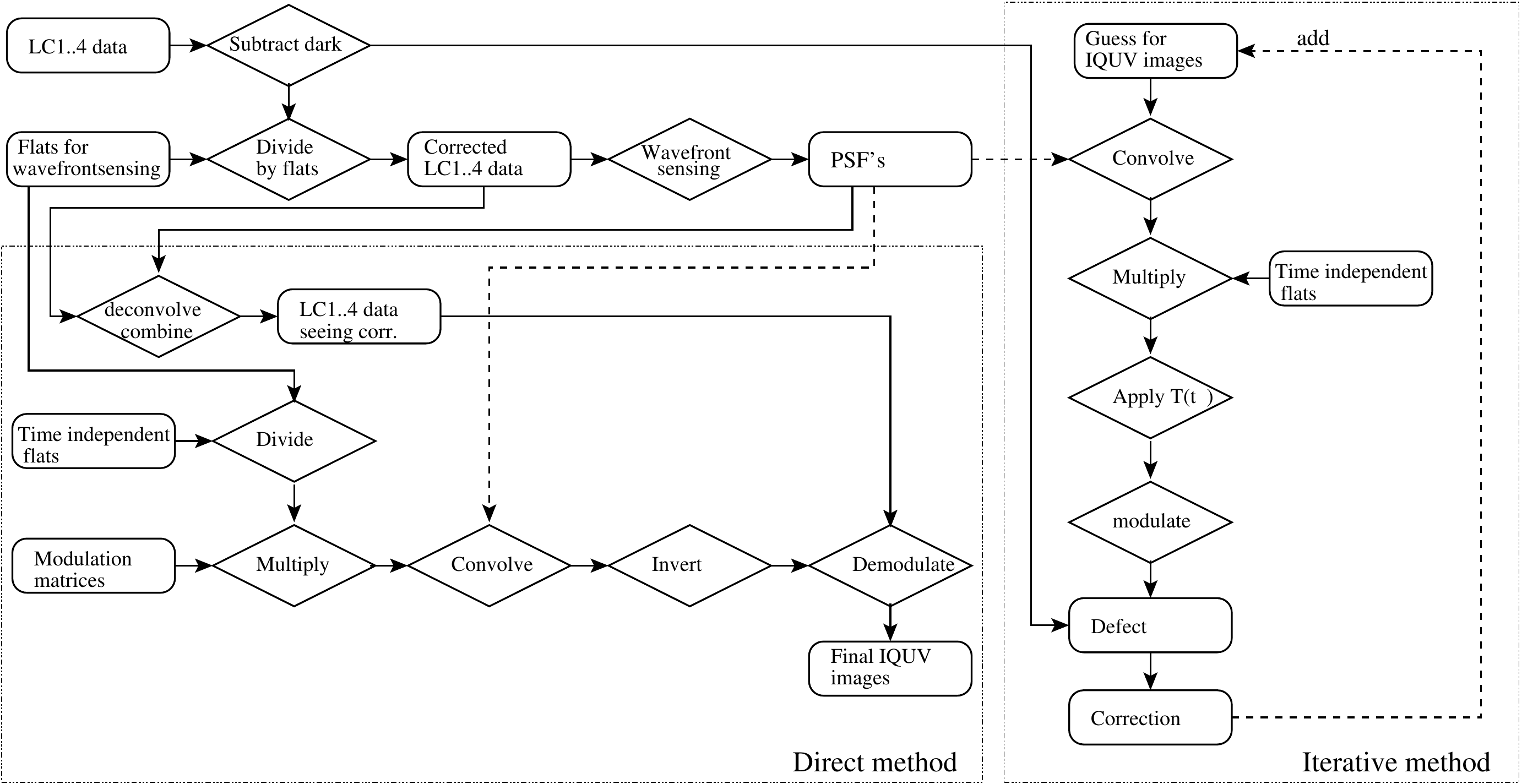}
\caption{Flowchart illustrating the direct and indirect method to calculate the final Stokes I, Q, U, and V images using image restoration.}
\label{fig:flowchart2}
\end{figure*}

An overview of both methods is presented in Fig.~\ref{fig:flowchart2}. For the direct method, flatfields for the time of the observations are determined as described in Sect.~\ref{sect:momfb_flats}, which are used for the image restoration process. These flatfields are then removed from the restored data and the true flatfield applied by multiplying the data with the ratio of the image restoration flatfield and the true flatfield, convolved with the average PSF.
The demodulation is then performed with the inverse of the modulation matrix convolved with the average PSF. The idea behind this is that each pixel in the restored image ''samples`` the whole PSF, and therefore the convolved calibration data should be applied.

An alternative method, which is simpler but computationally more expensive, is the iterative method, which will be described in more detail in a future paper we are preparing. This approach solves the entire forward problem, including all the calibrations and individual PSFs, and iteratively improves the incoming Stokes images to find the solution that fits the observed data best.

\section{Results}
\label{sect:results}

\subsection{Example: sunspot 6302}
\begin{figure*}[tbhp]
 \includegraphics[width=0.99\linewidth, trim=37 49 43.5 50, clip]{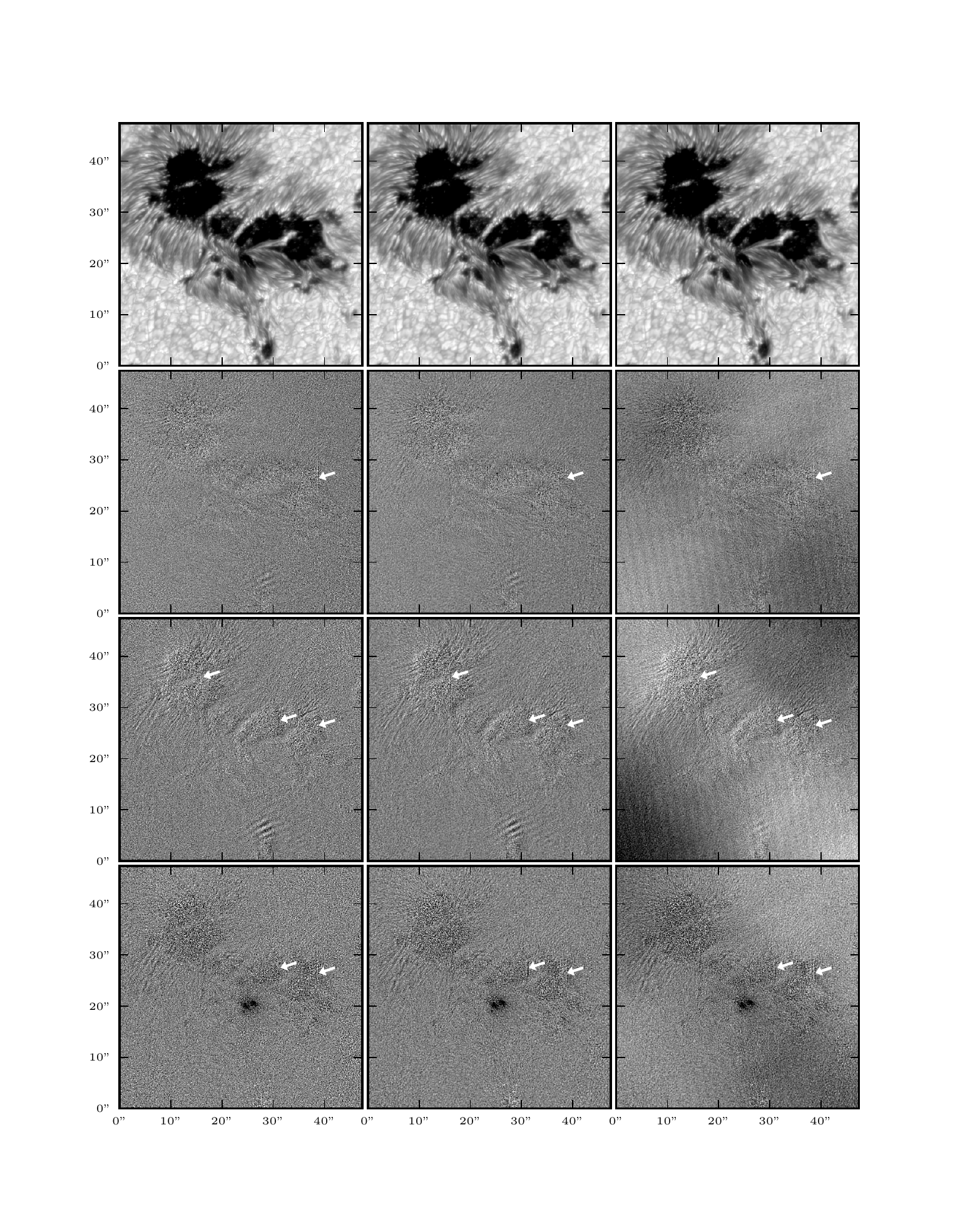}
 \caption{Continuum Stokes I, Q, U, and V images (from {\it top} to {\it bottom}) of the sunspot observed with the SST in the 6302.5 \AA\ \ion{Fe}{i} line on 27 October 2009. The images on the {\it left} were calibrated with the described calibration scheme and the iterative method, the images in the {\it middle} used the direct method instead, and the images on the {\it right} were calibrated by straightforwardly applying standard calibrations as listed in Sect.~\ref{sect:instr_calib} and using the direct method. The full range of the polarized images is $-$1.5\% to $+$1.5\%. The arrows indicate where sharp artifacts occur for some methods. Continuum offsets in the Q, U, and V images, most likely resulting from discrepancies in the telescope model, of up to $\sim$0.35\% have been subtracted. The small dark region visible in the Stokes V image is actual signal in a region with strong downflows (see text).}
 \label{fig:sunspot_cont}
\end{figure*}

As a test of our reduction method, we reduced a dataset from 27 October 2009, which has part of the $\delta$ sunspot AR11029 in the field of view (see Fig.~\ref{fig:sunspot_cont}) at a heliocentric angle of about 45$^{\circ}$ ($\mu\approx 0.7$). The data contain 11 equidistant wavelength points around the core of the \ion{Fe}{i} 6302.5 \AA\ line with a wavelength separation of 48 m\AA\ and a continuum point at +600 m\AA.

\begin{figure*}[tbhp]
\includegraphics[width=0.505\linewidth, trim=15 20 0 15, clip]{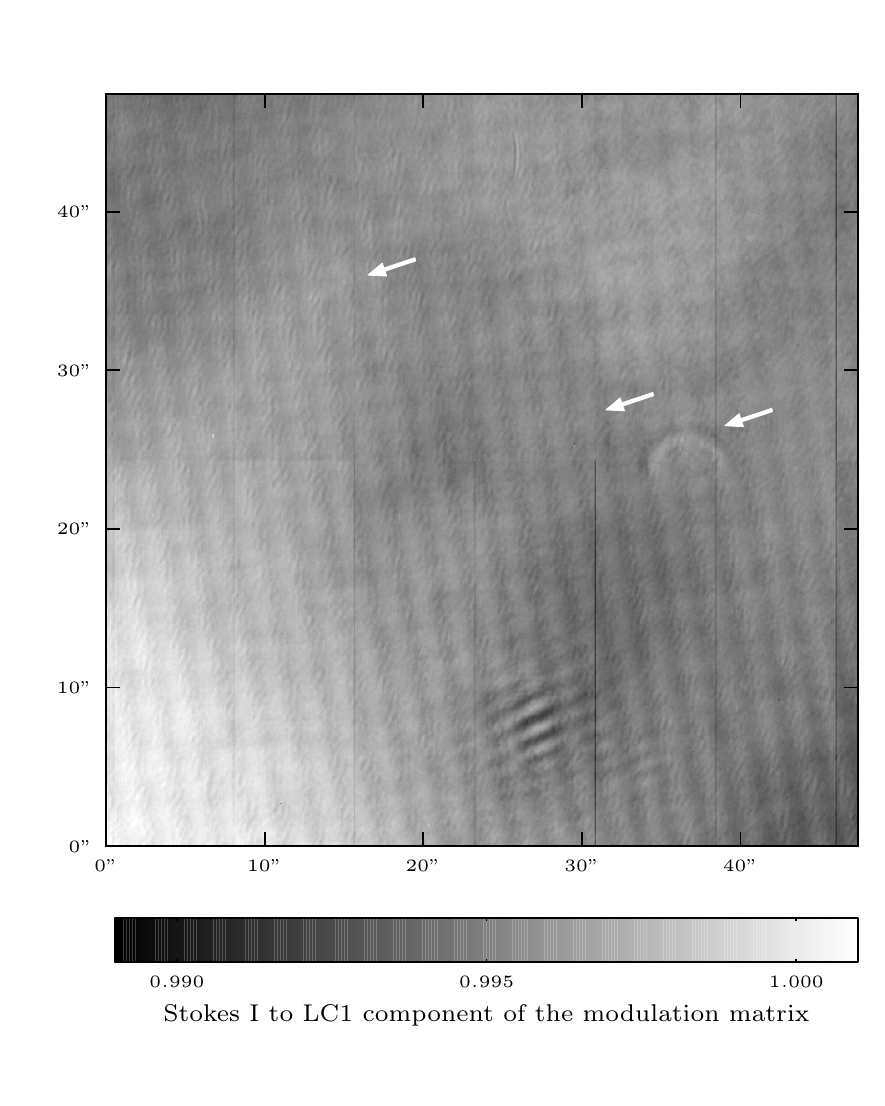}
\includegraphics[width=0.485\linewidth, trim=25 20 0 15, clip]{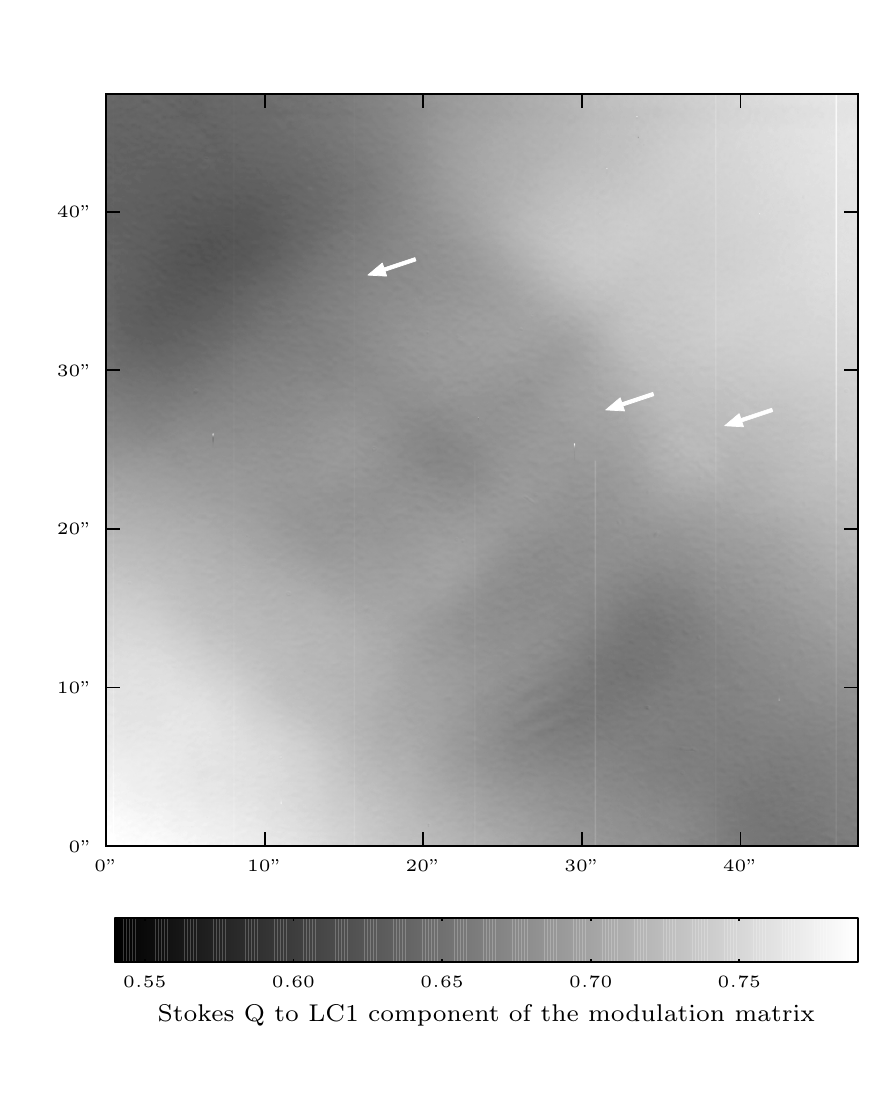}
\caption{Variations in modulation matrix elements over the FOV. The element describing the contribution of Stokes I to LC-state 1 is shown on the \textit{left} and the contribution of Stokes Q to LC-state 1 is shown on the \textit{right}.}
\label{fig:mmcomp}
\end{figure*}

\begin{figure*}[tbhp]
\includegraphics[width=\linewidth]{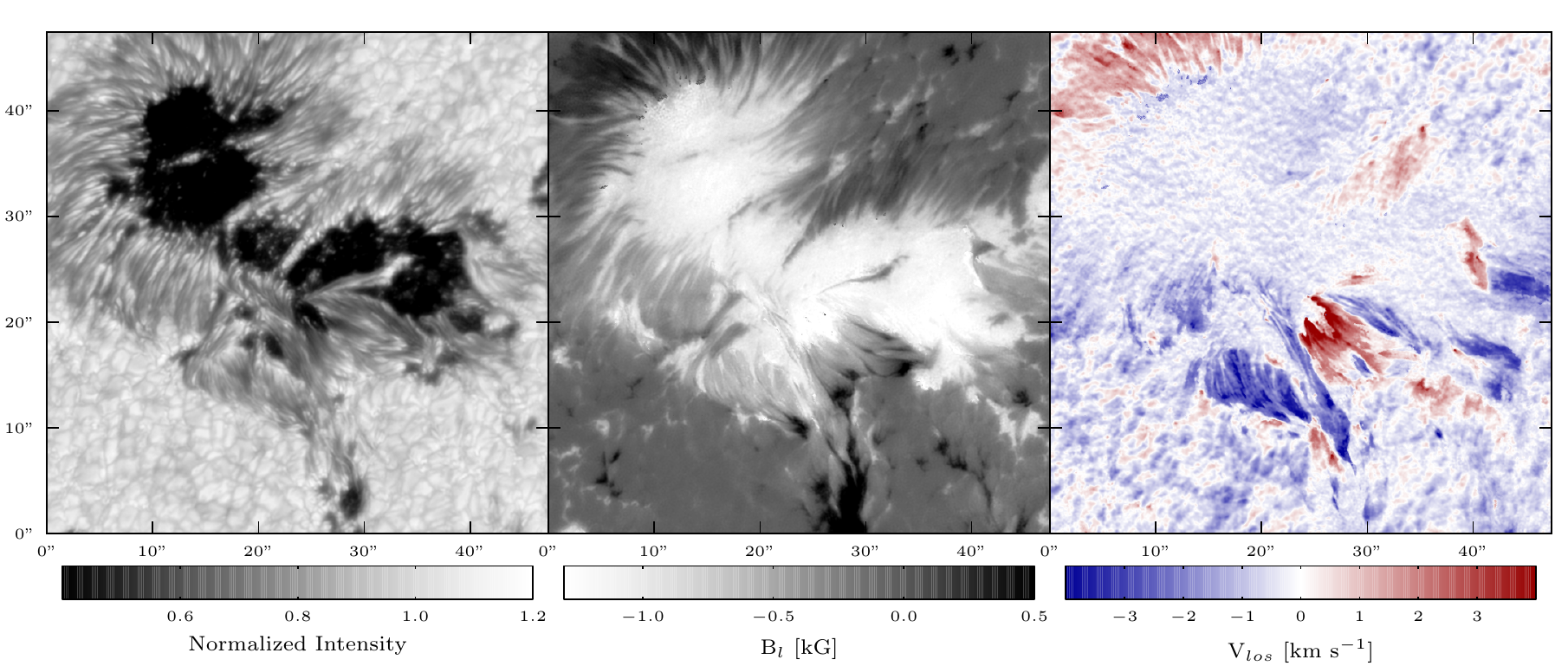}
\caption{Results of inversions of the data, showing normalized continuum intensity ({\it left}), the line-of-sight component of the magnetic field ({\it middle}), and the line-of-sight component of the velocity ({\it right}). The inversions reveal quite strong outflows (up to about 5 km\,s$^{-1}$) and strong downflows ($\sim$10 km\,s$^{-1}$). The limb is towards the upper right.}
\label{fig:inversions}
\end{figure*}

\subsection{Stokes parameters}
In Fig.~\ref{fig:sunspot_cont} we show the improvement in the Stokes parameters that is reached by using the scheme described in Sect.~\ref{sect:cal_scheme} as compared to straightforwardly applying standard calibrations as listed in Sect.~\ref{sect:instr_calib}. Polarimetric errors due to the changing telescope polarization that influences the flatfields are visible as saddle-shaped offsets, a vertical fringe-like pattern, and sharp lines in Stokes Q, U, and V. These errors almost completely disappear with the new calibration method.

The origin of these problems lies in the variations in the modulation matrix. In Fig.~\ref{fig:mmcomp} we show two representative examples of the 16 components of the modulation matrix: the contributions of Stokes I and Q to LC-state 1. As for all the LC-states, the Stokes I contribution shows a vertical fringe-like pattern over the whole FOV, some smaller scale patterns, likely related to dust, and sharp lines. The Stokes Q contribution is much smoother and mainly shows a saddle shape and some sharp lines, which is typical of the contributions of Stokes Q, U, and V to the LC-states. Because such patterns are present in the modulation and demodulation matrices, it is not possible to remove them both from the Stokes I, Q, U, and V images and from the corresponding modulated LC-states at the same time. Flats being chosen to make the Stokes images smooth, implies that the modulated images will not be smooth, and vice versa. Therefore our scheme uses two sets of flats, one to make the modulated images smooth for the wavefront sensing and one to properly flatfield the Stokes images.

The dark regions in the Stokes V map in Fig.~\ref{fig:sunspot_cont} are actual signals redshifted into the continuum wavelength due to strong downflows (see also Sect.~\ref{sect:inversions}). The small spot of fringes near the bottom of the Stokes Q, U, and V images is likely from a polarizing speck, e.g.\ dust, in the telescope, which is not accounted for in the telescope model. Outside these areas, the noise in the continuum point of the Stokes I normalized polarization images is $\sim$0.3\%.

\subsection{Inversion results}
\label{sect:inversions}
The data are inverted using the inversion code Nicole \citep{nicoleref}. For the inversions we use three nodes in temperature, one node (i.e.\ the parameter has a constant value) in line-of-sight velocity ($v_\mathrm{los}$), and one node for the line-of-sight magnetic field strength and the microturbulence. The average transmission profile (see Sect.~\ref{sect:real_gains}) is used to account for the finite spectral resolution of CRISP.

Results of the inversions are shown in Fig.~\ref{fig:inversions}. 
The shift in the velocities due to cavity errors of the FPI is corrected by subtracting a map of the cavity errors as determined in Sect.~\ref{sect:real_gains} convolved with the average PSF of the whole dataset.
The zero velocity reference was set to the average velocity in the umbra. Adopting this velocity reference, we find a convective blueshift as measured from the granulation in the FOV of 0.3 km\,s$^{-1}$, which is consistent with the theoretical value of 0.4 km\,s$^{-1}$ \citep{jaime_blueshift}.

Both the velocity and longitudinal magnetic field can reliably be determined from the data, showing fast ($\sim$5 km\,s$^{-1}$) outflows at the center-side penumbra and downflows (up to $\sim$10 km\,s$^{-1}$) on the intersection between the umbra and penumbra slightly below the middle of the FOV. These downflows will be discussed in more detail in a future paper.

\section{Conclusions}
\label{sect:conclusions}
We have presented a method for reducing full Stokes data obtained with CRISP at the SST, an instrument with a fast-tuning Fabry-P\'erot filter in a telecentric mount. Although some calibration issues are inherent to such an instrument used together with image restoration, most problems can be overcome by using the proper calibration procedures.

The impact on the flatfields of wavelength shifts over the FOV, owing to cavity errors, can be corrected for. This is important for preventing changes in the line shape caused by errors in the gain tables.
For instruments with the FPI in a collimated mount, the cavity errors occur in pupil space, and are therefore not a problem in this respect. These instruments do show radial wavelength shifts over the FOV that have to be accounted for \citep[][]{janssen:2006}.

Without a proper calibration scheme, the polarimetric imprint of the flatfielding can cause strong saddle-shaped offsets, fringes, and sharp lines from deviating pixel rows/columns that are enhanced by the deconvolution in the image restoration. By using the proposed calibration scheme these artifacts can almost be completely removed. Any remaining weak artifacts are most likely due to discrepancies in the SST telescope model, which is only accurate to a few tenths of a percent, and nonzero intrinsic polarization. The severity of the artifacts depends on the (variation in the) telescope polarization and the variations in the modulation matrix elements over the FOV. For telescopes that have low and/or constant intrinsic polarization and instrumentation whose modulation matrix elements are nearly constant over the FOV, such flatfielding problems will be much less severe.

Velocities, as determined from inversions of the data, can be corrected for cavity errors using a convolved version of the cavity map. However, 
the PSFs will vary from wavelength to wavelength resulting in a non equidistant wavelength sampling. Both the wavelength sampling and the variations in the width of the FPI transmission peak will vary from pixel to pixel, but can in principle be taken into account during the inversions. An intrinsic limitation is the use of non monochromatic input images for the image restoration, which results in deconvolution artifacts: mixing of information from different wavelengths through the image restoration procedure.

Modern developments in FPI instrumentation, such as rapid (kHz) modulation combined with charge shuffling (which may allow for demodulation before image restoration), could reduce calibration problems.
However, to reach the polarimetric and spectroscopic accuracy that the future large solar observatories EST and ATST are aiming for with the next generation of Fabry-P\'erot systems, advanced calibration techniques and optimized optical designs to deal with the issues discussed here will be required.

\acknowledgements{We would like to thank P.~S\"utterlin for taking the observations. This research project has been supported by a Marie Curie Early Stage Research Training Fellowship of the European Community's Sixth Framework Programme under contract number MEST-CT-2005-020395: The USO-SP International School for Solar Physics. The Swedish 1-m Solar Telescope is operated on the island of La Palma by the Institute for Solar Physics of the Royal Swedish Academy of Sciences in the Spanish Observatorio del Roque de los Muchachos of the Instituto de Astrof\'{i}sica de Canarias.}

\bibliography{./references}

\onecolumn
\begin{appendix}

\section{Determining the modulation matrix}
\label{appa}
When fitting the pixel-dependent system response matrix ${\cal M}$, it is important to realize 
that the incoming intensity $I$ is not known. A fit can therefore only be made on the
fractional polarization, which means that we need to normalize the data with the intensity. The
resulting data, lacking absolute transmission information, can only be used to determine the 
modulation matrix up to a constant factor $\lambda^{-1}$.
Moreover, the normalization of the data requires us to compute the intensity from
the data using the inverse of the unknown modulation matrix ${\cal M}$. It is therefore more 
natural to formulate the fit directly in terms of the demodulation matrix ${\cal M}^{-1}$:
\begin{equation}
{\cal M}^{-1}=\left(
\begin{array}{c c c c}
b_{11}&b_{12}&b_{13}&b_{14}\\
b_{21}&b_{22}&b_{23}&b_{24}\\
b_{31}&b_{32}&b_{33}&b_{34}\\
b_{41}&b_{42}&b_{43}&b_{44}\\
\end{array} \right). \nonumber
\end{equation}

The calibration data is taken using a linear polarizer followed by a $\lambda/4$-wave plate, which produces a known (relative) polarization, independent of polarization of the incoming light, by separately rotating the two plates. Given the calibration data $\vec{v}$ and the known (normalized) Stokes vector of the incoming light $\vec{S}$, we have for each 
calibration measurement point $n$ with unknown intensity $I_n$
\begin{equation}
b_{ij}v_{j,n} =S_{i,n} I_l=S_{i,n} \sum_{k=1}^4 b_{1k} v_{k,n}
\end{equation}
so that for a given guess-solution for ${\cal M}^{-1}$, the total accumulated fit error between the data and the 
guess matrix can be written as
\begin{equation}
\chi^2=\sum_{i=1}^4 \sum_n \left(  \sum_{j=1}^4 b_{ij}v_{j,n} - S_{i,n} \sum_{j=1}^4 b_{1j} v_{j,n}\right)^2.
\label{eq:chi2}
\end{equation}

Due to the use of normalized data, we have
\begin{equation}
\sum_{j=1}^4 b_{1j}=\lambda.
\label{eq:lambda}
\end{equation}
Substitution in Eq.~\ref{eq:chi2} eliminates the contribution to $\chi^2$ for $i=1$, as $S_{i,n}=1$. The 
minimum value of $\chi^2$ is then given by $\frac{\partial \chi^2}{\partial b_{kl}} = 0$, 
giving
\begin{align}
\frac{\partial}{\partial b_{kl}} \sum_{i=2}^4 \sum_n \left(  \sum_{j=1}^4 b_{ij}v_{j,n} - S_{i,n}\left[\lambda v_{1,n}+\sum_{j=2}^4b_{1j}(v_{j,n}-v_{1,n})\right]\right)^2 &= 0 \nonumber \\
\sum_{i=2}^4 \sum_n \frac{\partial}{\partial b_{kl}} \left(  \sum_{j=1}^4 b_{ij}v_{j,n} - S_{i,n}\left[\lambda v_{1,n}+\sum_{j=2}^4b_{1j}(v_{j,n}-v_{1,n})\right]\right)^2   &= 0.\nonumber
\end{align}
Applying the chain rule results in
\begin{equation}
2 \sum_{i=2}^4 \sum_n \left( \sum_{j=1}^4 b_{ij}v_{j,n} - S_{i,n}\left[\lambda v_{1,n}+\sum_{j=2}^4 b_{1j}(v_{j,n}-v_{1,n})\right]\right) \left(\sum_{j=1}^4 \delta_{ik}\delta_{jl}v_{j,n} - S_{i,n}\sum_{j=2}^4\delta_{1k}\delta_{jl}(v_{j,n}-v_{1,n})\right)=0, \nonumber
\end{equation}
where the last two sums over $j$ can be evaluated as
\begin{equation}
\sum_{i=2}^4 \sum_n \left( \sum_{j=1}^4 b_{ij}v_{j,n} - S_{i,n}\left[\lambda v_{1,n}+\sum_{j=2}^4b_{1j}(v_{j,n}-v_{1,n})\right]\right) \left(\delta_{ik} v_{l,n} - S_{i,n}\delta_{1k}(v_{l,n}-v_{1,n})\right)=0.
\end{equation}
This expression splits naturally into two separate cases:

For $k=1,\ l=2,3,4$:
\begin{align}
\sum_{i=2}^4 \sum_n & \left( \sum_{j=1}^4 b_{ij}v_{j,n} - S_{i,n}\left[\lambda v_{1,n}+\sum_{j=2}^4b_{1j}(v_{j,n}-v_{1,n})\right]\right) \left( -S_{i,n} (v_{l,n}-v_{1,n})\right)=0 \nonumber \\
\sum_{i=2}^4 \sum_n & \left( \sum_{j=1}^4 b_{ij}v_{j,n} - S_{i,n}\sum_{j=2}^4b_{1j}(v_{j,n}-v_{1,n})\right) S_{i,n}(v_{j,n}-v_{1,n}) = \lambda \sum_{i=2}^4 \sum_n S_{i,n}^2 v_{1,n} (v_{l,n}-v_{1,n})
\end{align}

and for $k>1,\ l=1,2,3,4$
\begin{align}
\sum_{i=2}^4 &\sum_n \left( \sum_{j=1}^4 b_{ij}v_{j,n} - S_{i,n}\left[\lambda v_{1,n}+\sum_{j=2}^4 b_{1j}(v_{j,n}-v_{1,n})\right]\right)\delta_{ik}v_{l,n}=0 \nonumber \\
&\sum_n \left( \sum_{j=1}^4 b_{kj}v_{j,n} - S_{k,n}\sum_{j=2}^4b_{1j}(v_{j,n}-v_{1,n})\right)v_{l,n}=\lambda \sum_n v_{l,n}S_{k,n}v_{1,n}
\end{align}

The solution of this system gives the least-squares fit to the data, which can be renormalized
by choosing an appropriate value for $\lambda$. The natural choice for $\lambda$ is $2$,
as the maximum intensity measured by one of the beams of the polarimeter is $\frac{1}{2}I$ for
unpolarized light. The one remaining element $b_{11}$ is constrained by Eq.~\ref{eq:lambda}.

\end{appendix}
\end{document}